\documentclass{iopjournal}

\usepackage{amsmath}

\begin{document}

\articletype{Paper} 

\title{SpinWaveToolkit: Python package for (semi-)analytical calculations in the field of spin-wave physics}

\author{Jan Klíma$^1$\orcid{0009-0002-1838-5546}, Ondřej Wojewoda$^{1,2,*}$\orcid{0000-0002-4276-520X}, Jakub Krčma$^3$\orcid{0009-0001-8647-1692}, Martin Hrtoň$^{1,3}$\orcid{0000-0002-3264-4025}, Dominik Pavelka$^1$\orcid{0009-0001-9253-5365}, Jakub Holobrádek$^1$\orcid{0000-0003-4632-2067}, and Michal Urbánek$^{1,3}$\orcid{0000-0003-0072-2073}}

\affil{$^1$CEITEC BUT, Brno University of Technology, Purkyňova 123, 612 00, Brno, Czech Republic}

\affil{$^2$Department of Materials Science and Engineering, Massachusetts Institute of Technology, 02139 Cambridge, Massachusetts, United States of America}

\affil{$^3$Institute of Physical Engineering, Brno University of Technology, Technick\'{a} 2, Brno, 616 69, Czech Republic}

\affil{$^*$Author to whom any correspondence should be addressed.}

\email{ondrej.wojewoda@vutbr.cz}


\begin{abstract}
We present  an open-source Python package, SpinWaveToolkit (SWT), for (semi-)analytical modeling of spin-wave dynamics in thin ferromagnetic films and exchange-coupled magnetic bilayers. SWT combines analytical models based on the Kalinikos-Slavin theory with a semi-analytical dynamic-matrix approach, enabling the calculation of dispersion relations, group velocities, decay lengths, mode profiles, and static equilibrium magnetization states. In addition, SWT implements a quantitative model of micro-focused Brillouin light scattering (BLS) that incorporates vectorial optical focusing, spin-wave Bloch functions, magneto-optical coupling, and Green-function propagation to simulate experimentally measured BLS spectra. The package is validated against finite-element dynamic-matrix simulations performed with TetraX for Damon-Eshbach, backward-volume, forward-volume, and oblique-field geometries, showing excellent agreement while reducing computation times by nearly two orders of magnitude in comparison to the numerical simulations. Thanks to the easiness of the use and fast calculation times, SWT can be used not only for exploratory mapping of the parameter space, but also for the fitting of the measured dispersion relations and related parameters. Thus, it provides a versatile and efficient framework for experiment design, interpretation, and parameter optimization for magnonics research.
\end{abstract}

\section{Introduction}
Almost all magnetic phenomena can be described by the Landau-Lifshitz-Gilbert (LLG) equation. This equation describes the damped precessional motion of the magnetization. Depending on the geometry and the set of interactions involved, a wide variety of physical phenomena can arise, such as spin waves, solitons, chaos, or resonances \cite{Lakshmanan_2011}. Due to the complicated nature of these interactions, LLG equation can be solved analytically only for a limited number of experimental geometries. This limitation has been addressed by the development of micromagnetic simulation tools, which allow one to define complex sample and experimental geometries \cite{abert_2019, Leliaert_2019}. However, since the LLG equation is a nonlinear partial integro-differential equation, these simulations usually require significant computational time. Substantial progress in this area has been achieved through advances in computational hardware, improved numerical algorithms, and the use of graphical processing units for selected steps of the calculation \cite{Vansteenkiste_2011, Vansteenkiste_2014, Donahue_OOMMF, moreels2024mumax, Chang_2011, Lepadatu_2020}.

In spin-wave research, one of the most important characteristics of a magnetic system is its dispersion relation. In order to extract the dispersion from space--time simulations, one must excite broadband magnetic dynamics and simulate the system over a relatively long time interval in order to achieve the desired frequency resolution \cite{Baker_2017, Venkat_2013}. This challenge was addressed by the dynamic-matrix approach, which allows the dispersion relation to be obtained directly \cite{korberFiniteelementDynamicmatrixApproach2021a, TetraX}. Within this framework, the LLG equation is not solved solely in the space-time domain, but along the direction of spin wave propagation in the frequency-wavevector domain. This significantly speeds up the calculation and allows fast design and optimization of various sample geometries and material parameters. Nevertheless, since this method still partially relies on solving the LLG equation in the spatial domain (typically in the waveguide cross-section), it remains computationally demanding for systematic exploration of large parameter spaces or for fitting the experimental results. 

Here, we address this challenge by introducing an open-source Python package, SpinWaveToolkit (SWT), which provides a collection of analytical and semi-analytical models for spin-wave physics, and is accompanied with online documentation and examples \cite{swt_docs}. These include dispersion relation calculations of the thin-film geometries under arbitrary external-field orientations based on the theory developed by Kalinikos and Slavin \cite{Kalinikos_1986, Kalinikos_1990}, semi-analytical evaluation of dispersion relations for in-plane and out-of-plane magnetized layers \cite{Tacchi_2019}, magnetic bilayers coupled via the Ruderman-Kittel-Kasuya-Yosida (RKKY) interaction \cite{gallardo_2019, wojewoda2024saf, wadge2024, glamsch2025magsaf}, and modeling of micro-focused Brillouin light scattering spectra \cite{wojewoda2024modeling, wojewoda2023observing, wojewoda2023phase}. Due to the easiness of the use and fast calculation times, this Python package can be used not only for the design of experiments by the exploratory mapping of the parameter space, but also for fitting the measured dispersion relation, group velocity, decay length or even Brillouin light scattering (BLS) spectra \cite{Vanatka2021, krcma_sciAdv, Turcan2021, Klima2024, flajsman_wideband_2022}.

\section{Modeling of thin ferromagnetic films and bilayers}
SWT contains two models for infinite thin ferromagnetic film: a fully analytical one based on the zeroth perturbation (class \texttt{SingleLayer}) and a semi-analytical approach which implements numerical calculation of the eigenvalues of the interaction matrix (class \texttt{SingleLayerNumeric}). The analytical model is slightly faster and can calculate dispersion relation for any in-plane and out-of-plane angle of the magnetization. However, depending on the parameters, this model can be inaccurate if the modes get into the vicinity of each other in the wavevector--frequency space, or in the region between dipolar and exchange dominated regimes. In contrast, the numerical model in its current state can be used only for an arbitrarily in-plane-magnetized and completely out-of-plane-magnetized layer. To address the rising interest in interfacially-coupled magnetic bilayers, e.g., synthetic antiferromagnets, SWT makes use of a two-macrospin model (class \texttt{DoubleLayerNumeric}) which semi-analytically solves the micromagnetic energy equilibrium and then solves the eigenvalues of the system matrix to retrieve the dispersion relation.

To handle the materials, SWT introduces a convenience class \texttt{Material}, which stores most of the magnetic material parameters relevant to the calculations (in SI units), such as saturation magnetization $M_\mathrm{s}$, gyromagnetic ratio $\gamma$, exchange stiffness constant $A_\mathrm{ex}$, Gilbert damping constant $\alpha$, and inhomogeneous broadening $\Delta H_0$, see Listing~1. The objects of this class can be then directly used in the dispersion models as input.

\begin{center}
    Listing 1: SWT import and material definition.
    \includegraphics[width=0.9\textwidth]{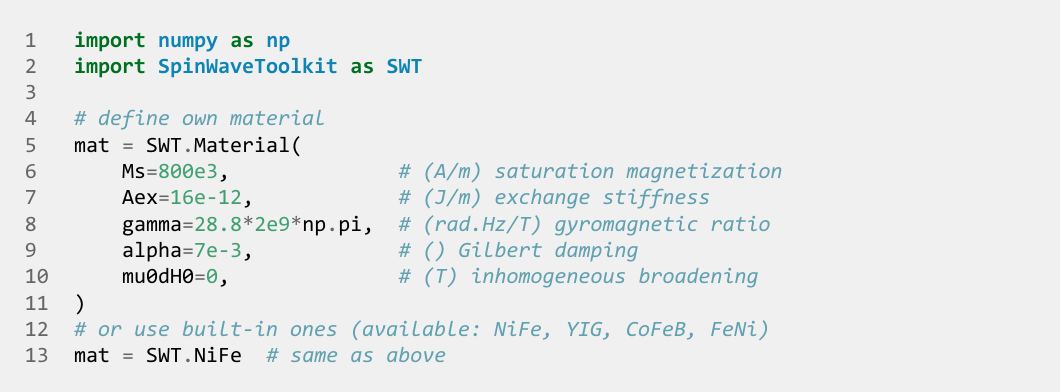}
\end{center}

\subsection{Fully analytical modeling -- \texttt{SingleLayer}}

For homogeneous thin films, the approach based on the zeroth-perturbation can be used to obtain the explicit expression for the spin-wave dispersion relation \cite{Kalinikos_1986}
\begin{equation}
\omega^2_n = \left( \omega_\mathrm{H} + \omega_\mathrm{M} l_\mathrm{ex}^2 k^2 \right)
\left( \omega_\mathrm{H} + \omega_\mathrm{M} l_\mathrm{ex}^2 k^2 + \omega_\mathrm{M} F_n \right),
\label{eq:SK_dispersion}
\end{equation}
where $\omega_\mathrm{H}=\mu_0 \gamma H_\mathrm{ext}$, $H_\mathrm{ext}$ is externally applied magnetic field, $\omega_\mathrm{M}=\mu_0 \gamma M_\mathrm{s}$, $l_\mathrm{ex}^{2} = 2 A_\mathrm{ex} \left(M_\mathrm{s}^2 \mu_0 \right)^{-1}$ is the squared exchange length, 
$k$ is the wavevector, and $n$ is the quantization number across film thickness. The function $F_n$ contains the magnetostatic contributions and depends on the propagation direction and film thickness. Our implementation allows for totally pinned, totally unpinned, and partially pinned (symmetric) boundary conditions. Moreover, one can define arbitrary direction of the propagation within the plane of the sample, calculate quantized modes along the film thickness ($n>0$) and define arbitrary demagnetizing and uniaxial anisotropy matrices. Even though the fully analytical implementation in SWT has a possibility to calculate spin-wave dispersion around the degeneracy points by using second perturbation correction, for simplicity we omit this in the following examples.

\begin{figure}[htb]
    \centering
    \includegraphics[width=\textwidth]{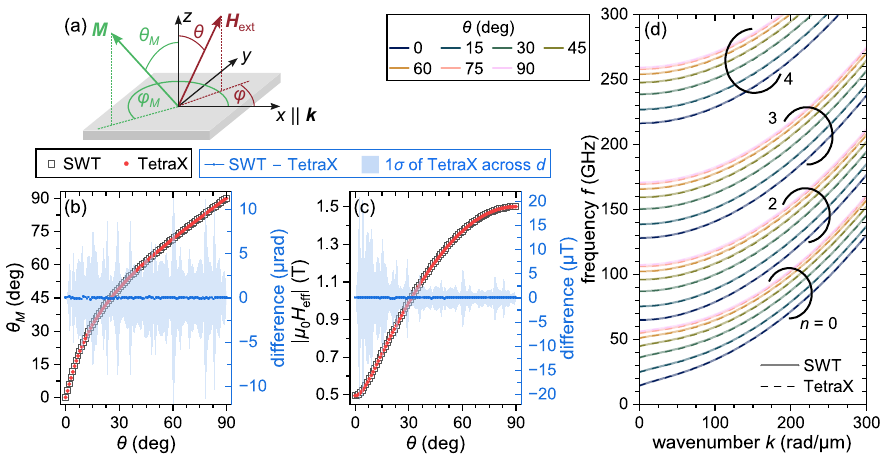}
    \caption{Dispersion calculation under oblique angle of external magnetic field. a, Sketch of the investigated geometry. b,\,c,~Magnetization polar angle $\theta_M$ (b) and effective field magnitude $|\mu_0 H_\mathrm{eff}|$ (c) results of minimization in SWT and TetraX and their difference. d,~Comparison of the dispersion relations under different polar angle of the external magnetic field $\theta$ for $n = 0, 2, 3, 4$ ($n=1$ omitted for clarity) and $\varphi=\pi/2$.}
    \label{fig2}
\end{figure}

In case of high-symmetry geometries, simple analytical formulas can be derived. However, in the case when the the external magnetic field is applied at an arbitrary angle, the calculations become more complicated as it is necessary to find the equilibrium position of the magnetization and the resulting effective magnetic field via energy minimization. The presented module offers a simple numerical energy minimization of a macrospin under external field, with arbitrary demagnetizing and uniaxial anisotropy tensors describing the sample. The total free energy density $\epsilon_\mathrm{tot}$ is then calculated as
\begin{equation}
    \epsilon_\mathrm{tot} = -\mu_0 M_\mathrm{s}\,\mathbf{m}\cdot\mathbf{H}_\mathrm{ext} + \frac{1}{2}\mu_0M_\mathrm{s}^2\,\mathbf{m}\hat{\mathbf{N}}_\mathrm{d}\mathbf{m} + \frac{1}{2}\mu_0M_\mathrm{s}^2\,\mathbf{m}\hat{\mathbf{N}}_\mathrm{uni}\mathbf{m},
\end{equation}
where $\mathbf{m}$ is the magnetization unit vector, $\hat{\mathbf{N}}_\mathrm{d}$ is the demagnetization tensor as a 3×3 matrix, and $\hat{\mathbf{N}}_\mathrm{uni}$ is the uniaxial anisotropy tensor [as a sum of 3×3 matrices of all (first-order) uniaxial anisotropies present in the system]. For simplicity, the model does not include any higher-order or multi-axial anisotropies in the current state. The evaluation of $\epsilon_\mathrm{tot}$ takes place in the periodic space $(\theta_M,\,\varphi_M)$, which relates to $\mathbf{m}$ simply by $\mathbf{m}=(\cos\varphi_M\sin\theta_M,\,\sin\varphi_M\sin\theta_M,\,\cos\theta_M)$, see Fig.~\ref{fig2}a.
Although this model is limited to spatially-invariant samples, it is sufficient to describe homogeneously-magnetized single thin films. This model is implemented as the \texttt{MacrospinEquilibrium} class and its output can be simply passed to the \texttt{SingleLayer} as is illustrated in Listing~2.

\begin{center}
    Listing 2: Example of use for the macrospin energy minimization and consequent input to the \texttt{SingleLayer} class, from which the dispersion relation for $n=0$ mode is calculated.
    \includegraphics[width=0.9\textwidth]{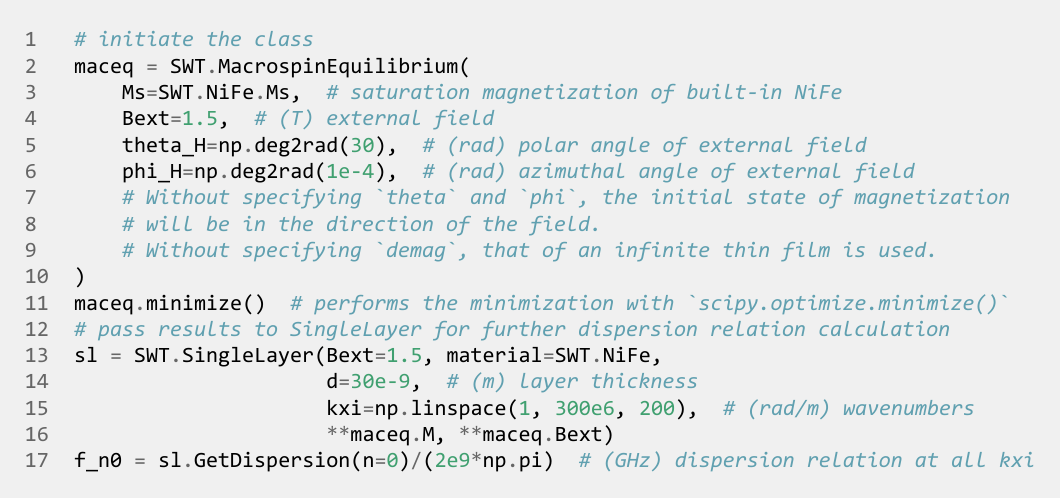}
\end{center}

To illustrate its relevance, we will compare the macrospin to a finite-element micromagnetic solver TetraX~\cite{TetraX}. The assumed geometry is sketched in Fig.~\ref{fig2}a.  In all calculations, we used standard parameters for a NiFe layer ($M_\mathrm{s} = 800\,\mathrm{kA\,m}^{-1}$, $A_\mathrm{ex}$ = 16\,pJ\,m$^{-1}$, $\gamma (2 \pi)^{-1} = 28.8\,$GHz\,T$^{-1}$ -- built in SWT as \texttt{SWT.NiFe})~\cite{korberFiniteelementDynamicmatrixApproach2021a, wojewoda2023observing, chumak2019, kalarickal2006}, 
with thickness of 30\,nm and external magnetic field magnitude set to $|\mu_0 H_\mathrm{ext}| = 1.5\,$T, while the polar (out-of-plane) angle of the external magnetic field $\theta$ is swept from 0 to 90 degrees. The energy minimization is used to find the equilibrium values of the in-plane and out-of-plane angles of magnetization (Fig.~\ref{fig2}b) and the internal magnetic field (Fig.~\ref{fig2}c). The same experimental geometry is also simulated in TetraX with 100 cells across the film thickness. The semi-analytical calculation is in perfect agreement in both, the equilibrium angle $\theta_M$ and in the calculated effective field $|\mu_0 H_\mathrm{eff}|$, with the finite-element solver (see blue lines in Fig.~\ref{fig1}b,\,c). While the thickness discretization might alter the parameters for a laterally confined or thicker layer, here their standard deviation $\sigma$ across film thickness does not exceed 10\,µrad and 20\,µT, respectively. Due to the spiky behavior of $\sigma$, we assume it comes mainly from numerical errors.

The acquired angle and internal field are subsequently used to calculate the spin-wave dispersion relation, see Fig.~\ref{fig2}d. Since the sample does not have any magneto-crystalline anisotropies ($\hat{\mathbf{N}}_\mathrm{uni} = 0$), the internal magnetic field is decreasing once the magnetization starts to tilt out of the plane of the sample, see Fig.~\ref{fig2}c. In the case of fully in-plane field ($\theta = 90\,$deg) the internal field is equal to the applied external field. In the other limiting case with external magnetic field pointing completely out-of-plane ($\theta=0\,$deg) the internal magnetic field is lowered by the demagnetizing field of the thin film [$\mu_0 H_\mathrm{eff} = \mu_0 \left( H_\mathrm{ext} - M_\mathrm{s} \right)$]. The decrease in the effective field causes the decrease of the spin-wave frequency $f=\omega\,(2\pi)^{-1}$ with fixed mode number, see Fig.~\ref{fig2}d.  

\subsection{Semi-analytical dynamic-matrix model -- \texttt{SingleLayerNumeric}}
To account for mode hybridization and avoided crossings, we employ a semi-analytical method based on the dynamic-matrix formulation \cite{Tacchi_2019}. In this approach, the linearized Landau-Lifshitz equation is written in matrix form
\begin{equation}
\omega\, \mathbf{m}_k = \hat{\mathbf{C}}_k \mathbf{m}_k,
\label{eq:dynamic_matrix}
\end{equation}
where $\mathbf{m}_k$ is a vector containing the mode amplitudes of the standing-wave basis and $\hat{\mathbf{C}}_k$ is the interaction matrix including exchange and dipolar coupling between thickness modes. The eigenvalues of $\hat{\mathbf{C}}_k$ yield the spin-wave dispersion, while the eigenvectors provide the mode profiles and relative mode weights.

Unlike the zeroth-order analytical approximation, this method fully captures intermode coupling and is therefore valid even in the presence of degeneracies and strong hybridization, at the cost of solving a numerical eigenvalue problem for each wavenumber $k$. In SWT, the $\hat{\mathbf{C}}_k$ matrix can be constructed for an arbitrary number of perpendicular standing spin-wave (PSSW) modes, which allows for a high-precision calculation even in relatively thick samples, where multiple modes are in close proximity. 

The syntax for this dynamic-matrix model, adapted in SWT as the \texttt{SingleLayerNumeric} class, is similar to all implemented dispersion models with some minor differences related to their nature (all explained in the respective documentation~\cite{swt_docs}). Listing 3 illustrates how to calculate the dispersion relation with \texttt{SingleLayerNumeric}.

\begin{center}
    Listing 3: Example of use for the \texttt{SingleLayerNumeric} class.
    \includegraphics[width=0.9\textwidth]{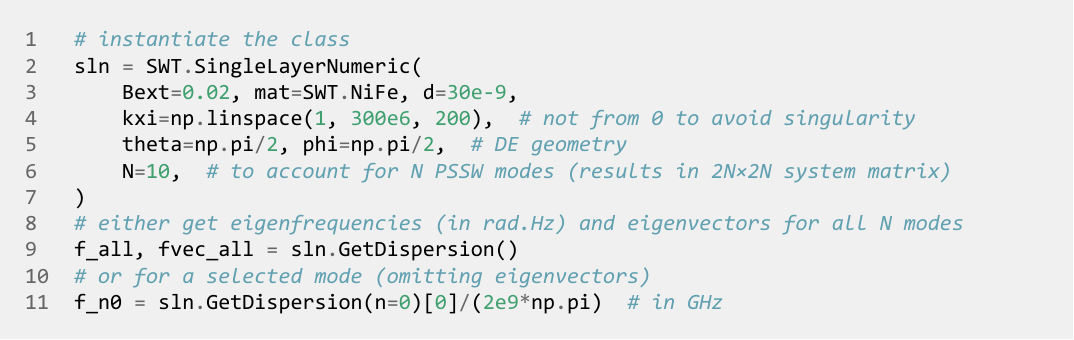}
\end{center}

\begin{figure}[htb]
    \centering
    \includegraphics[width=1\textwidth]{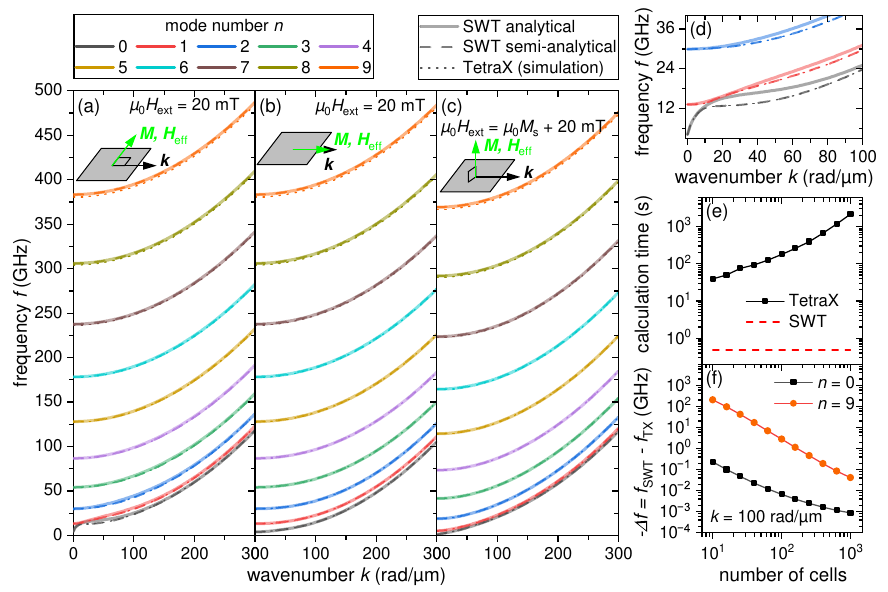}
    \caption{Dispersion relation calculation in three geometries with high symmetry and comparison with TetraX. a,\,b,\,c,~Dispersion relation of the first 10 spin-wave modes in Damon-Eshbach (a), backward-volume (b) and forward-volume (c) geometry. d,~Close-up to the mode hybridization visible at low frequencies and wavevectors in panel (a). e,~Comparison of the computation time between SWT (semi-analytical) and TetraX with different number of cells. It is given as the time needed to compute all mode frequencies at wavenumbers up to $n=10$ and $k=300\,$rad/µm, respectively. f, Comparison of the accuracy of TetraX with respect to \texttt{SingleLayerNumeric} for different number of cells for fundamental and $n=9$ modes.}
    \label{fig1}
\end{figure}

Fig.~\ref{fig1}a,\,b,\,c shows the dispersion relations of the first 10 thickness modes for Damon-Eshbach (DE), backward-volume (BV), and forward-volume (FV) geometries, respectively. The sample parameters are the same as in Fig.~\ref{fig2}, except that the magnetic field is changed to 20\,mT for in-plane geometries and $\sim$1.024\,T for FV ($\mu_0 M_\mathrm{s} + 20\,$mT). Furthermore, sample thickness was increased to 50\,nm, so that mode hybridizations become visible.
To validate this model, we again performed simulations in TetraX with the same material parameters and geometry. We used thin-film configuration and, if not stated otherwise, it was divided to 100\,cells in the out-of-plane direction to properly capture magnetization profiles in all 10 thickness modes. 

For both SWT models (analytical \texttt{SingleLayer} and semi-analytical \texttt{SingleLayerNumeric}), we observe agreement with the TetraX simulation across all calculated wavenumbers, except for regions where modes come close to each other in both, frequency and wavenumber. In this case, the zeroth perturbation theory (\texttt{SingleLayer}) is not valid anymore and we can observe significant differences, see Fig.~\ref{fig1}d. However, the semi-analytical model is still in a good agreement with the simulation result of TetraX.

The main advantage of the semi-analytical approach is the computation time. In TetraX, the computation time and precision are dependent on the number of cells, see Fig.~\ref{fig1}e,\,f. 
While the computation time increases almost linearly with the number of divisions~\cite{korberFiniteElementMonoAndMultiLayers2022}, the eigenmode frequency steadily approaches that from SWT, suggesting that the semi-analytical model gives very precise results in a fraction of time. This makes it ideal for rapid exploration of the parameter space and multi-dimensional calculations as will be shown later. On the other hand, the semi-analytical model is limited to very common but rather simple geometries. As noted in~\cite{korberFiniteElementMonoAndMultiLayers2022}, the large computation times of TetraX are mostly caused by the timely iterative diagonalization of the (sparse) dynamic matrix. 

\subsection{Magnetic bilayers -- \texttt{DoubleLayerNumeric}}

With the increasing interest of magnonics in multilayered systems, particularly interlayer exchange-coupled (IEC) structures, several theoretical frameworks have been developed to model and fit magnetization-reversal processes in such systems. These models range from simplified two-macrospin descriptions~\cite{gallardo_2019, glamsch2025magsaf} to approaches incorporating multiple discretization layers along the film thickness~\cite{wadge2024}. While more-general finite-element and finite-difference micromagnetic simulation tools provide high accuracy, they are computationally inefficient for systematic parameter fitting. SWT~implements the model proposed by Gallardo et al.~\cite{gallardo_2019} (as the \texttt{DoubleLayerNumeric} class). The two-macrospin approximation is particularly justified for ultrathin magnetic layers, where magnetization non-uniformity (twisting) along the thickness is negligible~\cite{glamsch2025magsaf}.

\begin{figure}[htb]
    \centering
    \includegraphics[width=\textwidth]{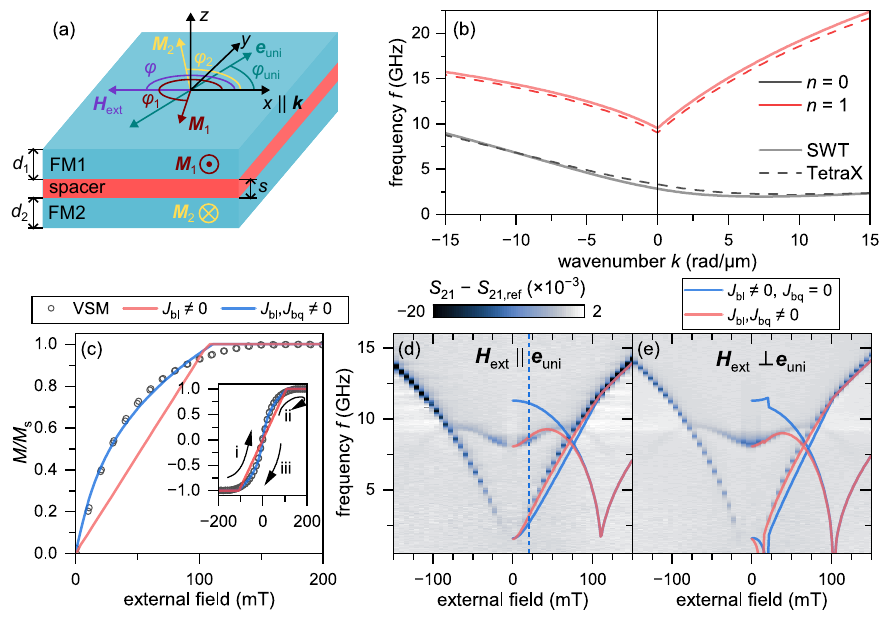}
    \caption{Static and dynamic calculations for a synthetic antiferromagnet (SAF) using the \texttt{DoubleLayerNumeric} class of SWT compared to TetraX and experiments for a CoFeB(15)/Ru(0.6)/CoFeB(15) sample. a,~Sketch of the investigated geometry. b,~Spin-wave dispersion relations at 20\,mT for an acoustic ($n=0$) and optical ($n=1$) mode. c,~Fits (solid lines) of an experimental hysteresis loop (circles) using SWT with and without the $J_\mathrm{bq}$ term. Inset shows the full loops and their acquisition direction (numbered arrows). Fixed parameters: $M_\mathrm{s}=1.175\,$MA\,m$^{-1}$, layer thicknesses, $K_\mathrm{u}=2\,$kJ\,m$^{-3}$, $\varphi-\varphi_\mathrm{uni}=\pi/2$. Fitted parameters: $J_\mathrm{bl}=-0.616(10)\,$mJ\,m$^{-2}$, $J_\mathrm{bq}=-0.163(3)\,$mJ\,m$^{-2}$ (blue line), $J_\mathrm{bl} = \left( -0.616 - 2\cdot0.163\right)\,$mJ\,m$^{-2}$, $J_\mathrm{bq}=0$ (red line). d,\,e,~Synthetic antiferromagnetic resonance experiment of the SAF sample half-overlaid with calculations from SWT for external field along~(d) and perpendicular~(e) to the uniaxial anisotropy axis. Vertical blue dashed line in d marks the settings used in b. Curves in b,\,d,\,e calculated with~\cite{wojewoda2024saf, Vanatka2021}: $M_\mathrm{s}=1\,$MA\,m$^{-1}$, $A_\mathrm{ex}=15\,$pJ\,m$^{-1}$, $\gamma(2\pi)^{-1}=30.5\,$GHz\,T$^{-1}$, $K_\mathrm{u}=2\,$kJ\,m$^{-3}$, $J_\mathrm{bl}=-0.6\,$mJ\,m$^{-2}$, and $J_\mathrm{bq}=-0.1\,$mJ\,m$^{-2}$ (unless stated otherwise).}
    \label{fig3}
\end{figure}

Similarly to the \texttt{SingleLayerNumeric} class, the core idea is to analytically calculate an interaction matrix $\hat{\mathbf{A}}$ and then numerically find its eigenvalues. But first, an equilibrium state of the free energy has to be found, in which this time all magnetic quantities are constrained to the plane of the films, as is illustrated in Fig.~\ref{fig3}a. The free energy density $\epsilon_\mathrm{tot}$ consists of dipolar, Zeeman, uniaxial anisotropy, and interlayer exchange contributions. The IEC part takes the form
\begin{equation}\label{eq:IEC}
    \epsilon_\mathrm{IEC} = -J_\mathrm{bl}\,\mathbf{m}_1\cdot\mathbf{m}_2 - J_\mathrm{bq}(\mathbf{m}_1\cdot\mathbf{m}_2)^2,
\end{equation}
where $J_\mathrm{bl}$ ($J_\mathrm{bq}$) denotes the bilinear (biquadratic) IEC constant. From Eq.~\eqref{eq:IEC}, it can be directly seen that the bilinear form of IEC favors an (anti)parallel orientation of $\mathbf{m}_{1}$, $\mathbf{m}_{2}$ (depending on the sign of $J_\mathrm{bl}$), while the biquadratic IEC favors either crossed ($J_\mathrm{bq}<0$) or collinear ($J_\mathrm{bl}>0$) orientation. Since all the parameters are constrained to the $xy$-plane, the evaluation of $\epsilon_\mathrm{tot}$ takes place in the $(\varphi_1,\,\varphi_2)$ space.

Once the equilibrium configuration has been determined, the corresponding eigenvalue problem is solved in order to obtain the spin-wave spectrum,
\begin{equation}
    \hat{\mathbf{A}}\mathbf{m} = \mathrm{i}\left(\frac{\omega}{\mu_0\gamma}\right)\mathbf{m},
\end{equation}
where $\mathbf{m}=(m_{1,\mathrm{IP}},\,m_{1,\mathrm{OOP}},\,m_{2,\mathrm{IP}},\,m_{2,\mathrm{OOP}})$ is the combined dynamic magnetization vector of the two layers and $\hat{\mathbf{A}}$ is the 4×4 interaction matrix incorporating all relevant magnetic contributions. This results in two spin-wave modes, which (in the saturated state) might be called, in analogy with phonons, an acoustic (in-phase precession) and optical (out-of-phase precession) mode. The syntax to obtain the dispersion relation from SWT is again similar to the previous case, as is illustrated by the example in Listing~4. Due to the interplay of the dipolar and exchange interactions between the two magnetic layers, the spin-wave dispersion relation in such a system is typically nonreciprocal in frequency, as is shown in Fig.~\ref{fig3}b. Simulating the same dispersion relation with TetraX revealed slight differences, which are likely attributable to the thickness discretization absent in the macrospin-based model.

\begin{center}
    Listing 4: Example of use for the \texttt{DoubleLayerNumeric} class.
    \includegraphics[width=0.9\textwidth]{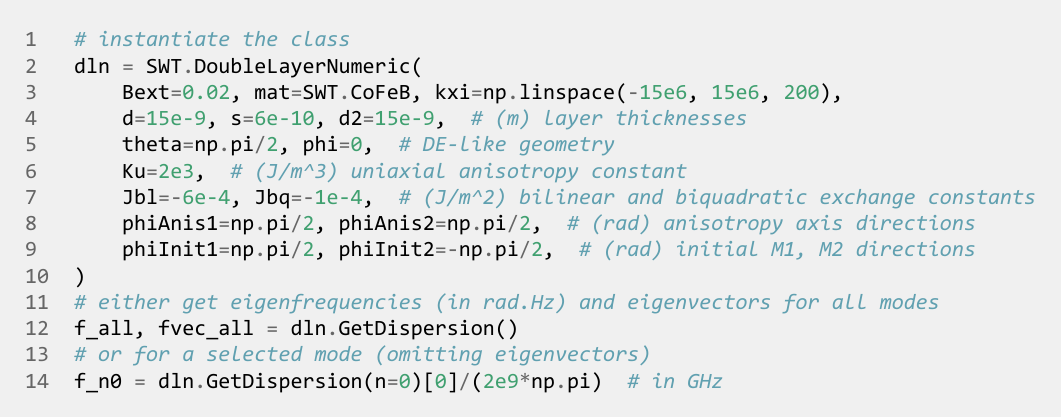}
\end{center}

\section{Applications of the models to more advanced problems}
In this section, we demonstrate that the computational efficiency and simplicity of the SWT framework enable its application to a wide range of practical tasks, including parameter fitting, optimization, and more advanced analyses that rely on the spin-wave dispersion relation.

\subsection{SAF hysteresis and VNA-FMR field sweep}\label{sec:safmr}
This example shows how to fit hysteresis loops and ferromagnetic resonance of synthetic antiferromagnets (SAFs) in order to obtain IEC constants and calculate non-reciprocal spin-wave dispersion. This approach was used in our recent work ~\cite{wojewoda2024saf} and yielded a good agreement with experimental observations and micromagnetic simulations.

One of the most common ways to estimate the IEC constants is fitting the hysteresis loop of the measured specimen~\cite{glamsch2025magsaf}. In Fig.~\ref{fig3}c, we present a hysteresis loop measured using a vibrating sample magnetometer (VSM, Quantum Design PPMS VersaLab), where the sample was a SAF consisting of the CoFeB(15)/Ru(0.6)/CoFeB(15) material stack on a silicon substrate (for exact composition, see~\cite{wojewoda2024saf}).

Fitting of the experimental data may be realized with standard Python workflows, such as the \verb|scipy.optimize.curve_fit()| function or the \texttt{lmfit} package, as well as in Matlab (via the Curve Fitting Toolbox) and OriginPro (via the Non-Linear Curve Fit dialogue window) thanks to their Python integration. This provides a practical advantage over existing IEC multilayer fitting tools that rely on dedicated graphical user interfaces, since it allows researchers to perform data preprocessing, modeling, fitting, and further analysis within a single software environment of their choice.

For the fitting, we used a custom function based on the \texttt{DoubleLayerNumeric} class, similar to the one shown in Listing~5. It takes an array of fields and fitting parameters as input, models the magnetization reversal consecutively in the supplied field values, and returns an array of normalized magnetization projected into the field direction. With this setup, we performed a fit of the VSM data by treating both, $J_\mathrm{bl}$ and $J_\mathrm{bq}$, as free parameters, see blue solid line in Fig.~\ref{fig3}c. To demonstrate the necessity to use both, bilinear and biquadratic components of IEC, we also calculated the hysteresis assuming only the bilinear coupling term of a magnitude giving the same saturation field, see red solid line in Fig.~\ref{fig3}c. This comparison clearly shows that assuming only the bilinear IEC term is not sufficient to correctly describe the experimental data. When biquadratic IEC is included, both, the initial susceptibility and curvature near saturation are increased and better agreement with the VSM data is reached. 

However, due to the relatively large thickness of the magnetic layers ($d>l_\mathrm{ex}$), the two-macrospin model fails to fully describe the experiment (e.g. the saturation field in Fig.~\ref{fig3}c), as some spin-twisting along the $z$ direction is present and to fully model the system, discretization would be needed. Therefore future SWT implementation could include some discretization of the FM layers, as was done e.g. in~\cite{wadge2024}, without significantly compromising the computational efficiency. Such an approach would allow the SWT to be applied to thicker multilayer systems and enable more accurate extraction of IEC constants, potentially including the intralayer exchange stiffness $A_\mathrm{ex}$.

Since $M_\mathrm{s}$ has large correlation coefficients with $J_\mathrm{bl}$ and $J_\mathrm{bq}$ in terms of the hysteresis loop fit, it is encouraged to first obtain the $M_\mathrm{s}$ by other means, e.g. from a VSM measurement in out-of-plane field, and then fit only the IEC constants with fixed $M_\mathrm{s}$.

\begin{center}
    Listing 5: A typical function used to fit the hysteresis loop of a synthetic antiferromagnet.
    \includegraphics[width=0.9\textwidth]{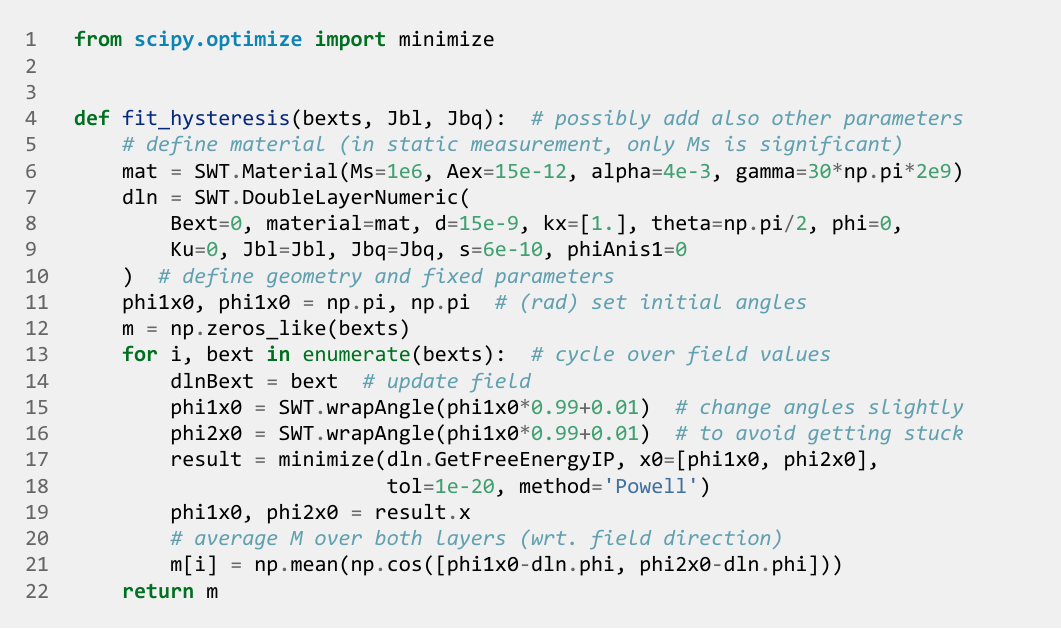}
\end{center}

In the next experiment, we measured microwave absorption in the same SAF sample. The microwave excitation and detection was facilitated by a vector network analyzer (Rohde \& Schwarz ZVA50), connected to a microwave printed circuit board (PCB), which was placed in a homogeneous magnetic field $\mathbf{H}_\mathrm{ext}$. The SAF was placed to a bend of the PCB signal trace, which enabled simultaneous excitation of the acoustic and optical modes~\cite{ishibashi2020}. The measured transmission signal, $S_{21}$ shown in Fig.~\ref{fig3}d,\,e was background-corrected by subtracting the reference $S_{21,\mathrm{ref}}$ as a field median at each frequency. The observed absorption signal corresponds to the synthetic antiferromagnetic resonance (SAFMR), i.e. spin waves with $k=0$. We measured with $\mathbf{H}_\mathrm{ext}$ along (Fig.~\ref{fig3}d) and perpendicular (Fig.~\ref{fig3}e) to the uniaxial anisotropy axis $\mathbf{e}_\mathrm{uni}$, which direction was previously determined by magneto-optical Kerr measurements \cite{wojewoda2024saf}.

In Fig.~\ref{fig3}d,\,e, we also present the calculated theoretical SAFMR spectra obtained using SWT. The acoustic mode is well described even without inclusion of the biquadratic interlayer exchange coupling, whereas the optical mode exhibits a qualitatively different field dependence when this term is neglected. In particular, the pronounced local minimum of the optical mode at zero field disappears when $J_\mathrm{bq}=0$, and its SAFMR is moved towards higher frequencies. Even though some local minimum is still present in the $\mathbf{H}_\mathrm{ext} \perp \mathbf{e}_\mathrm{uni}$ case due to the nonzero anisotropy strength (Fig.~\ref{fig3}e), it does not agree with the measured optical mode anymore.

\subsection{Two-dimensional dispersion}

So far, we have focused on dispersion relations along selected propagation directions. For in-plane magnetized thin films, however, the spin-wave dispersion $\omega(\mathbf{k})$ is strongly anisotropic, which can lead to e.g. non-collinearity of phase and group velocities, or global minima occurring away from the $\Gamma$~point ($\mathbf{k}\neq 0$).

\begin{figure}[htb]
    \centering
    \includegraphics[width=\textwidth]{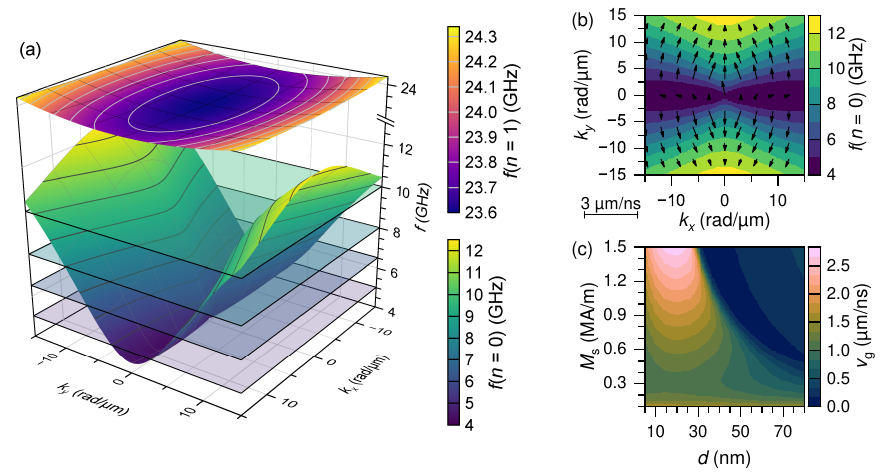}
    \caption{Two-dimensional dispersion relation and two-parameters optimization. a, The dispersion relation of fundamental and first perpendicular standing spin-wave mode for a 30\,nm thick NiFe layer (with the same parameters as in Fig.~\ref{fig2}) in 20\,mT in-plane field along the $x$ axis. The planes mark frequencies at which the Bloch function is shown in Fig.~\ref{fig5}a. b, Two-dimensional dispersion relation with calculated group velocity shown as arrows, which length represents group velocity magnitude $v_\mathrm{g}$ (see scale at left bottom). c, Optimization of $v_\mathrm{g}$ as a function of saturation magnetization $M_\mathrm{s}$ and thickness $d$ of the NiFe layer at $k=15$\,rad\,µm$^{-1}$. }
    \label{fig4}
\end{figure}

Because SWT enables rapid calculations for arbitrary in-plane propagation directions, the full two-dimensional dispersion surface $\omega(k_x,k_y)$ can be obtained, see Fig.~\ref{fig4}a. The material parameters are the same as in Fig.~\ref{fig2}, with the external magnetic field of 20\,mT applied in-plane along the $x$ axis. For the fundamental thickness mode ($n=0$), the dispersion exhibits pronounced anisotropy. There are two high-symmetry directions: the Damon-Eshbach geometry ($k_x=0$) and the backward-volume geometry ($k_y=0$). In contrast, the first perpendicular standing mode ($n=1$) is much less anisotropic, although weak anisotropy is still present.

The anisotropic dispersion generally implies that the group velocity $\mathbf{v}_\mathrm{g}$ is not collinear with the wavevector $\mathbf{k}$ (and hence with the phase-velocity direction). In two dimensions, the group velocity is obtained from the gradient of the dispersion,
\begin{equation}
\mathbf{v}_\mathrm{g}(\mathbf{k})=\nabla_{\mathbf{k}}\omega(\mathbf{k})
=\left(\frac{\partial \omega}{\partial k_x},\,\frac{\partial \omega}{\partial k_y}\right).
\end{equation}
Along the high-symmetry directions, $\mathbf{v}_\mathrm{g}$ becomes parallel or antiparallel to $\mathbf{k}$,
whereas at intermediate propagation angles it may strongly deviate due to the anisotropic curvature of the isofrequency contours (see Fig.~\ref{fig4}b). With the anisotropic dispersion relation, spin-wave caustics (or focusing directions) may occur at points of the isofrequency curve where the curvature vanishes. At these points, a finite range of wavevectors contributes to the same group-velocity direction, leading to an accumulation of energy flow along specific directions in real space~\cite{Wartelle2023}. In thin magnetic films, such conditions are typically realized in the vicinity of the Damon-Eshbach geometry, where $|\mathbf{v}_\mathrm{g}|$ reaches its largest values. Such two-dimensional calculations provide a natural starting point for modeling spin-wave propagation in thin films, including refraction and scattering from inhomogeneities.

\subsection{Optimization}

Beyond post-processing and data interpretation, SWT can be used prior to experiments or sample fabrication to identify optimal material and geometrical parameters. In many practical situations, one aims to maximize the spin-wave decay length or group velocity while simultaneously satisfying additional constraints, such as maintaining a sufficient frequency separation between the fundamental and first perpendicular standing spin-wave modes, or tuning the direction of caustic spin-wave propagation. Such optimization tasks can be efficiently addressed by systematic sweeps of the dispersion characteristics as a function of selected parameters.

As an illustrative example, we optimize the spin-wave group velocity in the Damon-Eshbach geometry at a fixed wavevector $k=15$\,rad\,µm$^{-1}$ as a function of film thickness $d$ and saturation magnetization~$M_\mathrm{s}$. The resulting two-parameter optimization landscape is shown in Fig.~\ref{fig4}c. Apart from $d$ and $M_\mathrm{s}$, all parameters are identical to those used in Fig.~\ref{fig1}a. The calculated landscape exhibits a complex, non-monotonic behavior, reflecting the interplay between exchange, dipolar interactions, and mode hybridization. In the following, we qualitatively describe the main observed trends.

For low saturation magnetization ($M_\mathrm{s} < 0.15\,\mathrm{MA}\,\mathrm{m}^{-1}$), the group velocity remains small and positive, and is nearly independent of the film thickness. In this regime, the exchange length is large and the spin waves are already exchange dominated at the chosen wavevector, leading to weak sensitivity to the sample thickness. 

At higher saturation magnetization and for thin films ($d < 30\,\mathrm{nm}$), the group velocity initially increases with thickness. This behavior originates from the growing contribution of the dipolar interactions, which drives a gradual crossover from the exchange-dominated to the dipolar-dominated regime and steepens the dispersion relation.

For thicker films ($d > 30\,\mathrm{nm}$), the group velocity decreases again. This reduction is caused by the onset of mode hybridization between thickness modes, which leads to a flattening of the dispersion relation near the selected wavevector. This example highlights how SWT can be used to identify optimal parameter regions while simultaneously revealing the physical mechanisms that limit spin-wave performance in the studied systems.

\subsection{Modeling of the micro-focused Brillouin light scattering signal}
Apart from dispersion relation calculation, SWT can be also used to model micro-focused Brillouin light scattering (BLS) spectra~\cite{wojewoda2024modeling, Benaziz2025}. All implemented models for calculation of the dispersion relation can be utilized for BLS signal calculation. Moreover, the Bloch function (density of spin-wave states in wavevector-frequency space) can also be obtained by other means, such as micromagnetic simulations, and SWT will be only used to calculate the BLS signal. All related functions are part of the sub-module \texttt{SWT.bls}.

In the first step, the incident electric field needs to be calculated. For this, we utilize the approach developed by Richards and Wolf \cite{richards1959electromagnetic, wolf1959electromagnetic, novotny2012}. It leads to an expression of the focal field in cylindrical coordinates $\mathbf{E}(\rho, \varphi, z)$ (explicitly given in~\cite{wojewoda2024modeling}) which evaluation requires several numerical integration steps.

\begin{figure}[htb]
    \centering
    \includegraphics[width=1\textwidth]{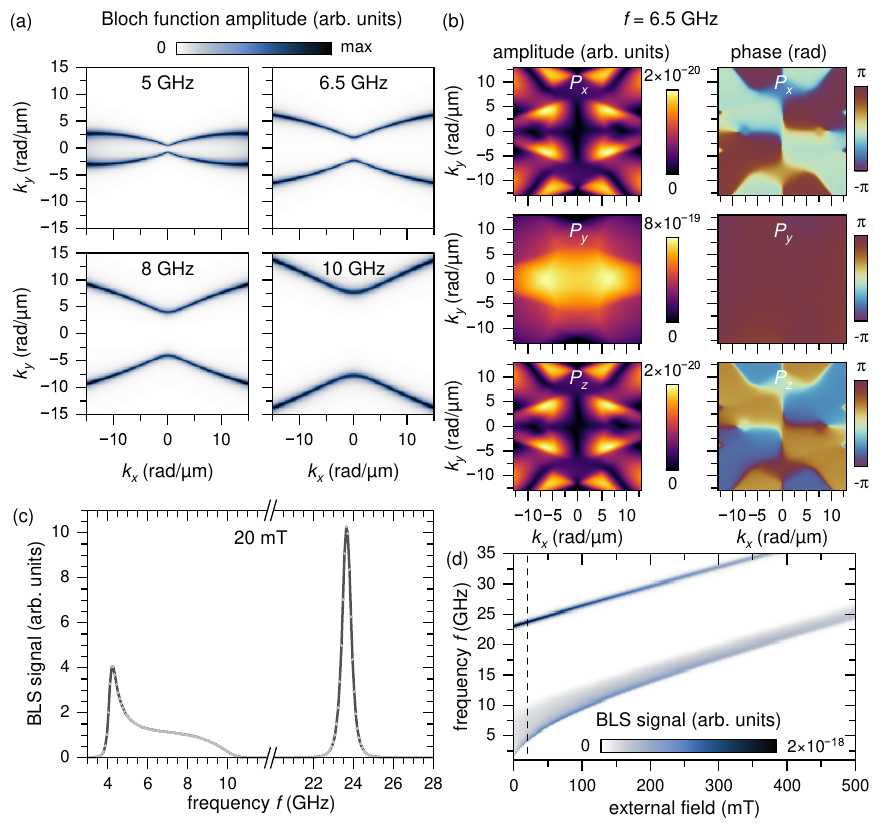}
    \caption{Calculation of the micro-focused BLS signal. a, Spin-wave density of states (Bloch function) for four selected frequencies. Material and experiment parameters are same as in Fig.~\ref{fig4}a. b, Polarization vector magnitude and phase at 6.5\,GHz. c, Calculated BLS signal for external magnetic field of 20\,mT. d, Calculated BLS spectra in different external magnetic fields. The values shown in panel c are marked by the vertical black dashed line.}
    \label{fig5}
\end{figure}

In the following step, the Bloch function has to be calculated. In order to allow for calculation in all models implemented in SWT, we follow this phenomenological equation
\begin{equation}\label{eq:bloch}
            \mathcal{D} (\omega, \mathbf{k}) \propto  \sqrt{2}\left(\exp \left(\frac{\hbar\omega-\mu}{k_\mathrm{B} T}\right)-1\right)^{-\frac{1}{2}} \sum_{i} \frac{1}{\left( \omega - \omega_i(\mathbf{k}) \right)^2 + \left( \frac{2}{\tau_i (\mathbf{k})} \right)^2},
\end{equation}
where $\mu$ is the magnon chemical potential, $T$ is the temperature, $k_\mathrm{B}$ is the Boltzmann constant, and $\hbar$ is the reduced Planck constant. The sum in~\eqref{eq:bloch} goes over all relevant PSSW modes with the corresponding dispersion relation $\omega_i(\mathbf{k})$ and lifetime $\tau_i(\mathbf{k})$. The results of this calculation for a material with the same parameters as in Fig.~\ref{fig4}a are shown for four selected frequencies in Fig.~\ref{fig5}a. The Bloch function is composed of individual spin-wave resonances located at the position of the dispersion relation with linewidth defined by their respective lifetimes (eigen states are shown in Fig.~\ref{fig4}a with marked constant frequency planes).

The next step is calculation of the induced polarization $\mathbf{P}$ in the magnetic layer
\begin{equation}
\label{Eq:Pqvec}
\mathbf{P}(\omega,\mathbf{k}_\mathrm{p},z)= \hat{{\chi}} (\omega_{\mathrm{m} },\mathbf{k}_{\mathrm{m} },z) \mathbf{E} (\omega - \omega_{\mathrm{m} },\mathbf{k}_\mathrm{p} - \mathbf{k}_\mathrm{m} ,z) ,
\end{equation}
where $\omega$ is the frequency of the induced polarization, $\mathbf{k}_\mathrm{p}$ is its in-plane (parallel to the magnetic layer) wave vector, while $\omega_\mathrm{m}$ and $\mathbf{k}_\mathrm{m}$ are their magnon counterparts. The electric susceptibility $\hat{{\chi}}$ is given by the magneto-optical coupling mechanism
\begin{equation}
\hat{{\chi}} = 
\left(
\begin{array}{ccc}
 0 & i M_z Q & -i M_y Q \\
 -i M_z Q & 0 & i M_x Q \\
 i M_y Q & -i M_x Q & 0 \\
\end{array} \right) ,
\end{equation}
where the dynamic components of magnetization are calculated from the Bloch function.

The calculated polarization for frequency $f=6.5\,$GHz is shown in Fig.~\ref{fig5}b. The component perpendicular to the original polarization ($P_\mathrm{y}$) is approximately one order of magnitude larger than $P_\mathrm{x}$ and $P_\mathrm{z}$. Even though the polarization states with wavevector larger than that of free-space light ($k_0 \approx 12\,\mathrm{rad}\,$µm$^{-1}$ at 532\,nm wavelength) can be induced, the radiation they generate is trapped in the near-field, as there is no mechanism for these high-momentum states to out-couple to propagating free-space light and reach the detector.

In order to get the electric field intensity incident on the detector, we utilized the Green's function~$\hat{\mathbf{G}}$ formalism. The BLS signal is then given by
\begin{equation}
\sigma (\omega_{\mathrm{m} } )  = \! \int \!\! \mathrm{d}^2 r_{\parallel} \int \!\! \mathrm{d}^2 k_{\mathrm{m} } \bigg\vert \int\limits_{k \leq k_{0} \mathrm{NA} } \!\!\!\!\!\!\! \mathrm{d}^2 k \, e^{\mathrm{i} \mathbf{k} \cdot \mathbf{r}_{\parallel}} \int \mathrm{d}^2 k^{\prime} \hat{\mathbf{G}}
\, \mathbf{P}\bigg\vert^{2}.
\label{eq:BLSthermal}
\end{equation}

The resulting BLS spectrum is shown in Fig.~\ref{fig5}c. The dominant narrow peak at low frequencies is caused by the low group velocity, and thus high density of states, around the ferromagnetic resonance frequency. The gradual decrease of intensity with increasing frequency is caused primarily by the decrease of the detection sensitivity towards higher wavevectors. 

One of the typical magnonic experiments is measurement of the BLS spectra in varying external magnetic field. We modeled such experiment for our system, see Fig.~\ref{fig5}d. The peak at lower frequencies comes from the fundamental mode, while the peak at higher frequencies originates from the $n=1$ mode.

\section{Conclusion}
In this paper, we presented the Python package SpinWaveToolkit (SWT) for fast (semi-)analytical calculations of spin-wave and magnetization dynamics. SWT provides models for calculating dispersion relations, group velocities, lifetimes, decay lengths, spin-wave mode profiles, and magnetic equilibrium states in thin films and exchange-coupled bilayers -- everything accompanied with extensive documentation \cite{swt_docs}. For thin films, both fully analytical and semi-analytical dynamic-matrix approaches are implemented, while bilayers are treated within the semi-analytical framework. In addition, SWT includes a complete module for modeling micro-focused Brillouin light scattering (BLS) signals.

We validated all implemented models by comparison with TetraX simulations for Damon-Eshbach, backward-volume, and forward-volume geometries, as well as for thin films under oblique magnetic fields. In all cases, SWT reproduces the results of simulations, including mode hybridization effects, while being approximately two orders of magnitude faster. The ability to efficiently explore material and experimental parameter spaces, model dispersion characteristics, and  micro-focused BLS spectra makes SWT a powerful tool for the design, interpretation, and optimization of modern magnonics experiments and devices.

\ack{We gratefully akcnowledge the colormaps used within this work~\cite{crameri2020,crameri2023zenodo}.}

\funding{O.W. was supported by Horizon Europe - MSCA grant agreement No. 101211677, project FeriMag. This research was supported by the project No. CZ.02.01.01/00/22 008/0004594 (TERAFIT). We acknowledge CzechNanoLab Research Infrastructure supported by MEYS CR (LM2023051).}

\roles{CRediT: Conceptualization: OW; Data curation: JKl; Formal Analysis: JKl; Funding acquisition: OW, MU; Investigation: JKl, OW, JH; Resources: JH; Software: JKl, OW, MH, JKr, DP; Supervision: OW, MU; Validation: ; Visualization: JKl, OW; Writing – original draft: JKl, OW; Writing – review \& editing: MH, JKr, DP, JH, MU.}

\data{All data and code used to generate the presented figures are available in the Zenodo repository~\cite{swt_zenodo}.}


\bibliographystyle{iopart-num}
\bibliography{bibliography}

\end{document}